\documentclass[epj-spec]{svjour}
\usepackage{graphics}
\usepackage{amssymb}
\usepackage{color}
\newtheorem{thm}{Theorem}
\newcommand{\half}{\frac{1}{2}}
\begin{document}
\title{Complementarity  in atomic  (finite-level quantum)  systems: an
information-theoretic approach}
\author{R. Srikanth\inst{1} \and Subhashish Banerjee\inst{2}
}                     
\institute{ Poornaprajna Institute of Scientific Research, Sadashiva
Nagar, Bangalore- 560 080, India
\and 
Raman Research Institute, Sadashiva Nagar,
Bangalore - 560 080, India
}
\date{Received: date / Revised version: date}
%

\abstract{
We develop an information theoretic interpretation of the number-phase
complementarity  in  atomic  systems,  where  phase is  treated  as  a
continuous  positive  operator valued  measure  (POVM).  The  relevant
uncertainty  principle is  obtained  as an  upper  bound on  a sum  of
knowledge  of  these  two   observables  for  the  case  of  two-level
systems. A  tighter bound  characterizing the uncertainty  relation is
obtained numerically  in terms of  a weighted knowledge  sum involving
these variables.   We point out that complementarity  in these systems
departs from  mutual unbiasededness in two  signalificant ways: first,
the maximum knowledge  of a POVM variable is  less than log(dimension)
bits;  second,  surprisingly,  for  higher  dimensional  systems,  the
unbiasedness  may  not be  mutual  but  unidirectional  in that  phase
remains unbiased  with respect to  number states, but not  vice versa.
Finally, we study the  effect of non-dissipative and dissipative noise
on these complementary variables for a single-qubit system.}
%

%
\maketitle

\section{Introduction}
Two  observables  $A$  and  $B$  of  a  $d$-level  system  are  called
complementary  if  knowledge of  the  measured  value  of $A$  implies
maximal  uncertainty of  the measured  value  of $B$,  and vice  versa
\cite{kraus,mu88}.   Complementarity is  an aspect  of  the Heisenberg
uncertainty  principle, which  says  that for  any  state $\psi$,  the
probability  distributions obtained  by measuring  $A$ and  $B$ cannot
both  be   arbitrarily  peaked  if   $A$  and  $B$   are  sufficiently
non-commuting.  Heisenberg  uncertainty is traditionally  expressed by
the relation
\begin{equation}
\triangle_\psi A
\triangle_\psi B \ge \frac{1}{2} |\langle [A,B]\rangle_\psi|,
\label{eq:hu}
\end{equation}
where  $(\triangle_\psi  A)^2  =  \langle A^2\rangle_\psi  -  (\langle
A\rangle_\psi)^2$.   However, this  representation  of the  Heisenberg
uncertainty relation has the disadvantage  that the right hand side of
Eq.   (\ref{eq:hu})  is   not  a  fixed  lower  bound   but  is  state
dependent. For example,  if $\psi$ is an eigenstate  of $A$, then both
$\triangle_\psi  A$ and  the  right hand  side  of Eq.   (\ref{eq:hu})
vanish, so that  no restriction is imposed on  the uncertainty in $B$.
To improve  this situation, an information  theoretic (or ``entropic")
version of  the Heisenberg uncertainty relationship  has been proposed
\cite{kraus,mu88,deu83},   which   relies   on  Shannon   entropy   of
measurement outcomes as a measure of uncertainty \cite{nc00,delg}.  An
application of  this idea to  obtain an entropic  uncertainty relation
for oscillator systems in the Pegg-Barnett scheme \cite{pb89} has been
made  in Ref.  \cite{abe},   and for  entropic uncertainty
relations   among   more   than   two  complementary   variables,   in
Ref. \cite{wiwe}. 

Given two  observables $A \equiv  \sum_a a|a\rangle\langle a|$  and $B
\equiv  \sum_b b|b\rangle\langle  b|$,  let the  entropy generated  by
measuring  $A$  or  $B$  on   a  state  $|\psi\rangle$  be  given  by,
respectively,   $H(A)$   and   $H(B)$.   The   information   theoretic
representation  of the  Heisenberg uncertainty  principle  states that
$H(A) + H(B)  \ge 2\log\left(\frac{1}{f(A,B)}\right)$, where $f(A,B) =
\max_{a,b}|\langle a|b\rangle|$, and  $H(\cdot)$ is the Shannon binary
entropy.   We  note that  $f(A,B)  \ge  d^{-1/2}$,  where $d$  is  the
(finite) dimension of the system.  A pair of observables, $A$ and $B$,
for which  $f(A,B)=d^{-1/2}$ are said to form  mutually unbiased bases
(MUB) \cite{ii81,dur05}.  Thus, any  $|a\rangle$ is an equal amplitude
superposition   in   the  basis   $\{|b\rangle\}$   and  vice   versa.
Conventionally,  two Hermitian  observables  are called  complementary
only if they are mutually unbiased.  Given a mutually unbiased pair of
Hermitian  observables,  $A$   and  $B$,  the  Heisenberg  uncertainty
relation takes the form
\begin{equation}
H(A) + H(B) \ge \log d.
\label{eq:hu0}
\end{equation}
A  further  advantage  of  the  entropic version  of  the  uncertainty
principle  over  (\ref{eq:hu})  is  that  unlike  the  latter,  it  is
insensitive  to  eigenvalue  relabeling,   and  depends  only  on  the
probability distribution obtained  by measuring $A$ or $B$  on a given
state \cite{deu83}.

Even the  information theoretic representation  (\ref{eq:hu0}) may not
in general  be suitable  if $A$  or $B$ is  not discrete,  because the
continuous  analog   of  $H(A)$,  which  is   $H_c(p)  \equiv  -\int_x
dx~p(x)\log[p(x)]$, is not positive definite,  as can be seen from the
case where  the probability distribution  is given by $p(x)=2$  for $x
\in [0,\frac{1}{2}]$  and $p(x)=0$ for $x  \in (\frac{1}{2},1]$, where
we find  $H_c(p)=-\log2$.  It is well possible  that this pathological
behaviour does not afflict classes of physical states of interest.  In
particular,  we  verified   this  in  the  case  of   the  {\em  phase
distribution} of two- and  four-level atomic systems.  However, we are
not aware that  this is generically true. In  any case, this potential
problem  can be  generally overcome  if the  uncertainty  principle is
expressed    in    terms   of    relative    entropy   (also    called
Kullb{\"a}ck-Leibler    divergence,   which   is    always   positive)
\cite{kl51},  instead of Shannon  entropy.  An  example of  where this
finds application  would be when one  of the observables,  say $A$, is
bounded,  and  its conjugate  $B$  is  described  not as  a  Hermitian
operator but as  a continuous-valued POVM.  An instance  of this kind,
considered  below in  detail, is  the number  and phase  of  an atomic
system. Here we show that the relative entropic definition can be used
to express  complementarity of number  and phase, where the  notion of
complementarity is  extended to accomodate POVMs.  We  thus makes this
intuitive  notion  more  concrete.   Here  the  `number'
variable 
is analogous to energy in oscillator systems (in the sense of having
discrete eigenvalues with fixed difference between consecutive values)
and amplitude of light field (eg., a laser, in the sense of being
conjugate to a phase variable).
We  note that  recourse to
relative entropy  is not  necessary for a  POVM of  discrete variables
\cite{mas07},  since Shannon  entropy is  well defined  in  this case.

The  quantum description  of  phases \cite{pp98}  has  a long  history
\cite{pb89,pad27,sg64,cn68,ssw90,scr93}.   Pegg   and  Barnett
\cite{pb89},  following  Dirac   \cite{pad27},  carried  out  a  polar
decomposition  of the  annihilation operator  and defined  a Hermitian
phase  operator  in  a  finite-dimensional Hilbert  space.   In  their
scheme, the expectation  value of a function of  the phase operator is
first carried out in a  finite-dimensional Hilbert space, and then the
dimension  is taken  to  the limit  of  infinity. However,  it is  not
possible to interpret this expectation  value as that of a function of
a Hermitian  phase operator  in an infinite-dimensional  Hilbert space
\cite{ssw91,mh91}.  To circumvent  this problem, the concept of phase
distribution   for    the   quantum   phase    has   been   introduced
\cite{ssw91,as92}.    In   this  scheme,   one   associates  a   phase
distribution to a  given state such that the average  of a function of
the phase operator in the state, computed with the phase distribution,
reproduces the results of Pegg and Barnett.

An interesting  question to ask  is how mutually  unbiased observables
behave in the  presence of noise.  Intuitively, one  would expect that
the uncertainty or entropy of each observable should be non-decreasing
under the  effect of noise.  However,  this is not  generally true, as
seen for example  in the case of a  quantum deleter \cite{qdele,sr07},
where uncertainty  in the computational  basis vanishes asymptotically
during a qubit's dissipative interaction  with a vacuum bath.  Here we
study  number   and  phase  of   atomic  systems  subjected   to  both
non-dissipative and dissipative  noise.  Noise can be thought  of as a
manifestation  of  an  open  system  effect  \cite{bp02}.   The  total
Hamiltonian is  $H = H_S +  H_R + H_{SR}$  , where $S$ stands  for the
system,  $R$  for the  reservoir  and  $SR$  for the  system-reservoir
interaction.  The  evolution of  the system of  interest $S$  (in this
case the atomic  system) is studied taking into  account the effect of
its  environment $R$, through  the $SR$  interaction term,  making the
resulting  dynamics  non-unitary.   The  open system  effects  can  be
broadly  classified into  non-dissipative, corresponding  to  the case
where $[H_S, H_{SR}] = 0$ resulting in decoherence without dissipation
or dissipative, corresponding to the case where $[H_S, H_{SR}] \neq 0$
resulting in decoherence along with dissipation \cite{bg07}.

A class of observables that  may be measured repeatedly with arbitrary
precision,  with the  influence of  the measurement  apparatus  on the
system being confined strictly to the conjugate observables, is called
quantum  non-demolition  (QND)   or  back-action  evasive  observables
\cite{bvt80,bk92,wm94,zu84}.   Such  a  measurement  scheme  was
originally introduced in the context of the detection of gravitational
waves  \cite{ct80,bo96}.   The  non-dissipative open  system  effect
described above would be a  QND effect.  Since they describe dephasing
without dissipation,  a study of  phase diffusion in such  a situation
would  be important  from  the  context of  a  number of  experimental
situations.  A study of the quantum phase diffusion in a number of QND
systems  was  carried  out   in  Ref.   \cite{sb06}  using  the  phase
distribution  approach.  In  Ref.   \cite{sr07}, the  above study  was
extended to include the effect of dissipation on phase diffusion. This
would  be  under the  rubric  of  a  dissipative open  system  effect,
described above.

In this paper  we study  three broad,  related problems:
first,  we  formulate  a  novel  characterization  of  the  Heisenberg
uncertainty relationship in terms of Kullb\"ack-Leibler divergence (or
relative entropy).  Second,  we motivate it by applying  it to a study
of  complementarity in an  angular momentum  system, which  involves a
continuous   variable  POVM;   lastly,  we   study  the   behavior  of
complementary   variables   when    subjected   to   dissipative   and
non-dissipative (purely dephasing) noise.  

The plan of the paper  is as follows.  In Section \ref{sec:phasdistr},
we introduce  the concept  of phase distribution  in an  atomic system
which  would  be used  subsequently.   In  Section \ref{sec:qinf},  we
motivate  and  develop  an  information  theoretic  representation  of
complementarity as applied to a  two-level atomic system, with a brief
discussion  of  a  four-level  atomic  system.  Since  any  system  of
interest  would, inevitably,  be  surrounded by  an environment  which
would effect  its dynamics,  it is of  relevance to discuss  the above
ideas of complementarity in the context of open quantum systems. We do
this  in  Section   \ref{sec:open}  by  recapitulating  relevant  work
\cite{sr07,bg07,sb06,gp,sqgen} on  open quantum systems,  of relevance
here. Section \ref{sec:openphase}  deals with the non-dissipative open
system    effect,   described   by    the   phase    damping   channel
\cite{nc00,bg07,sb06,gp}, and  Section \ref{sec:opengen} discusses the
dissipative open system effect,  described by the squeezed generalized
amplitude damping  channel \cite{gp,sqgen}.  The reason  for the above
terminologies  is the connection  of the  dynamics generated  by these
processes  with the  noise  effects pertinent  to quantum  information
\cite{gp}.   For  completeness,  we  relegate some  technical  details
pertaining to these  noisy channels to Appendix A  and B, 
where  the  physical  processes  underlying these  channels  are  also
briefly discussed.   In Section \ref{sec:concl} we make
our  conclusions and  discuss some  open questions  coming out  of our
work.

\section{Phase distribution \label{sec:phasdistr}}

It is not possible to interpret the expectation value of a function of
the phase operator, in the Pegg and Barnett scheme \cite{pb89}, as the
expectation value  of a function of  a Hermitian phase  operator in an
infinite-dimensional Hilbert space \cite{ssw91,mh91}.  This motivates
the  introduction of  the  phase distribution  for oscillator  systems
\cite{ssw91,as92}.  Interestingly,  the concept of  phase distribution
can also  be extended  to atomic systems  \cite{as96}, which  we study
here .  The phase  distribution ${\cal P}(\phi)$, $\phi$ being related
to the phase of the dipole moment of the system, is given by
\begin{equation} 
{\cal   P}(\phi)  =   {2j+1  \over   4  \pi}   \int_{0}^{\pi}  d\theta
\sin(\theta) Q(\theta, \phi), \label{2a.4}
\end{equation}
where  ${\cal   P}(\phi)>  0$  and  is  normalized   to  unity,  i.e.,
$\int_{0}^{2\pi} d\phi {\cal  P}(\phi) = 1$. In the  above, $j$ is the
angular  momentum  of  the  atomic  system.  The  quantity  $\phi$  is
important in  the context of atomic coherences  and the interferometry
based  on  such  coherences  \cite{as96}. Here  $Q(\theta,  \phi)$  is
defined as
\begin{equation}
Q(\theta, \phi)  = \langle \theta, \phi|\rho^s|  \theta, \phi \rangle,
\label{2a.5}
\end{equation}
where  $|\theta,   \phi  \rangle$  are  the   atomic  coherent  states
\cite{mr78,ap90}  given by an expansion over  the Wigner-Dicke states
\cite{at72},  which are  the simultaneous  eigenstates of  the angular
momentum operators $J^2$ and $J_Z$, as
\begin{equation}
|\theta, \phi \rangle = \sum\limits_{m= -j}^j 
\left(\matrix{2j \cr j + m}\right)^{1 \over 2} 
(\sin(\theta / 2))^{j+m} 
(\cos(\theta / 2))^{j-m} |j, m \rangle e^{-i(j + m) \phi}. 
\label{2a.6} 
\end{equation} 
It can be shown that the angular momentum operators 
$J_\xi, J_\eta$ and $J_\zeta$ (obtained by rotating the operators $J_x,
J_y$ and $J_z$ through an angle $\theta$ about an axis $\hat{n}
= (\sin\phi, -\cos\phi,0)$), being mutually non-commuting,
obey an uncertainty relationship of the type
$\langle J_\xi^2 \rangle
\langle J_\eta^2 \rangle \ge \frac{1}{4}\langle J_\zeta^2 \rangle$.
Atomic coherent states 
(obtained by rotating the Wigner-Dicke states via
$\theta$ and $\phi$ as above)
are precisely those states that saturate this bound, from which the name
is derived \cite{as96}. For two level systems,
they exhaust all  pure states, whereas for larger  dimensions, this is
no  longer  true. Using  Eq.   (\ref{2a.5})  in  Eq.
(\ref{2a.4}), with insertions  of partitions of unity in  terms of the
Wigner-Dicke states,  we can write the phase  distribution function as
\cite{sb06}
\begin{eqnarray} 
{\cal  P}(\phi) &=&  {2j+1 \over  4 \pi}  \int_{0}^{\pi}  d\theta \sin
\theta \sum\limits_{n,m=  -j}^{j} \langle  \theta, \phi |j,  n \rangle
\langle j,  n| \rho^s (t)| j, m \rangle \langle
j, m| \theta, \phi \rangle. \label{2a.7}
\end{eqnarray} 
The  phase  distribution ${\cal  P}(\phi)$,  taking  into account  the
environmental effects, have been studied  in detail for QND as well as
dissipative  systems in  \cite{sb06,sr07}  for physically  interesting
initial conditions  of the system $S$, i.e.,  (a). Wigner-Dicke state,
(b). atomic coherent state and (c). atomic squeezed state.

\section{Information theoretic representation of complementarity
\label{sec:qinf}}

The relative  entropy associated  with a discrete  distribution $f(j)$
with respect to a distribution $g(j)$ defined over the same index set,
is given by
\begin{equation}
S(f||g) = \sum_j f(j)\log\left(\frac{f(j)}{g(j)}\right).
\label{eq:re0}
\end{equation}
It can  be thought of as  a measure of  `distance' of 
distribution $f$ from  distribution $g$ in
that  $S(f||g)  \ge 0$,  where the equality holds
 if  and only  if  $f(j)=g(j)$
\cite{nc00}. Consider random variable $F$ with probability distribution
$f$. We will define $R(F)$ as the relative entropy of $f$ with
respect  to the uniform  distribution $\frac{1}{d}$, i.e.,
\begin{equation}
R(F) \equiv R[f(j)] = \sum_j f(j)\log(df(j)).
\label{eq:rf}
\end{equation}
As  a measure  of  distance  from a  uniform  distribution, which  has
maximal  entropy, $R(F)$  can  be  interpreted as  a  measure of  {\em
knowledge}, as  against uncertainty, of the  random variable described
by  distribution  $f$.   The  following  theorem  re-casts  Heisenberg
uncertainty principle in terms of relative entropy.

\begin{thm}
Given  two mutually unbiased  Hermitian observables  $A$ and  $B$, the
uncertainty relation (\ref{eq:hu0}) is equivalent to
\begin{equation}
R(A) + R(B) \le \log d,
\label{eq:ra}
\end{equation}
where $d$ is the (finite) dimension of the system.
\end{thm}
{\bf Proof.} Let the distribution obtained by measuring $A$ and $B$ on
a given state be, respectively, $\{p_j\}$ and $\{q_k\}$.  The l.h.s is
given by
\begin{eqnarray}
S\left(A||\frac{1}{d}\right) + S\left(B||\frac{1}{d}\right) &=&
\sum_j p_j \log( dp_j) + 
\sum_k q_k \log( dq_k) \nonumber \\
&=& -[H(A) + H(B)] + 2\log d \label{eq:sub} \\
&\le& -2\log\left(\frac{1}{f(A,B)}\right) + 2\log d.
\end{eqnarray}
This  is  the  general  result  for any  two  non-commuting  Hermitian
observables.    If   $A$  and   $B$   are   mutually  unbiased,   then
$f(A,B)=d^{-\half}$, and the theorem follows.  
It follows from the concavity of $H$, and thus from the
convexity of $R$, that the inequality Eq. (\ref{eq:ra}) derived for
pure states holds also for mixed states. 
\hfill $\blacksquare$
\bigskip

Physically,  Eq. (\ref{eq:ra})  expresses the  fact  that simultaneous
knowledge of  $A$ and $B$  is bounded above  by $\log d$.  This  is in
contrast to inequality (\ref{eq:hu0}), which is bounded below, being a
statement  on  the  sum  of  ignorances or  uncertainties.   Both  are
equivalent  ways   of  expressing   the  fact  that   the  probability
distributions obtained  by measuring $A$ and $B$  on several identical
copies of a given state  cannot both peak simultaneously.  

In terms of $R$, two Hermitian  observables $A$ and $B$ of a $d$-level
system are  called mutually unbiased  if the maximal knowledge  of the
measured  value  of $A$,  given  by  $\log  d$ bits,  implies  minimal
knowledge of  the measured  value of  $B$, given by  0 bits,  and vice
versa.   In  anticipation of  the  introduction  of  POVMs instead  of
Hermitian  observables,  we will  find  it  convenient  to weaken  the
definition of mutual  unbiasedness and call two variables  $A$ and $B$
(of which  one or both of them  may be a POVM)  as {\em quasi-mutually
unbiased}  if the  maximal  knowledge  of the  measured  value of  $A$
implies  minimal knowledge  of the  measured  value of  $B$, and  vice
versa.   The maximum  knowledge no  longer  being $\log  d$ bits,  but
lesser, the  pair $A$  and $B$ may  be called  quasi-mutually unbiased
bases (quasi-MUB's), an extension of  the concept of MUB from the case
of orthonormal bases to that of non-orthonormal bases.

If two observables  are not mutually unbiased, then  $\log d$ does not
bound from above the knowledge sum $R_T \equiv R(A) + R(B)$, and there
exist states such that the corresponding sum satisfies $R_T > \log d$.
Intuitively, this is  because in the case of  two observables that are
not mutually unbiased, knowledge  of the two observables pertaining to
a  given state  may simultaneously  peak.  For  example,  consider the
qubit  observables $\sigma_z$  and ${\bf  n}\cdot{\bf \sigma}$  in the
Hilbert     space     $\mathbb{C}^2$,     where     ${\bf     n}     =
(\sin\theta,0,\cos\theta)$ and  ${\bf \sigma}$ is the  vector of Pauli
matrices.  It can be seen using Eq. (\ref{eq:sub}) that any eigenstate
of ${\bf n}\cdot{\bf \sigma}$ corresponds  to the knowledge sum $R_T =
2 -  H(\cos^2(\theta/2))$.  This sum  is greater than one,  except for
$\theta=\pi/2$, which corresponds  to the mutually unbiased observable
$\sigma_x$.

Eq.  (\ref{eq:re0}) has a  natural extension  to the  continuous case,
given by
\begin{equation}
S(f||g) = \int dp~ f(p)\log\left(\frac{f(p)}{g(p)}\right).
\label{eq:re1}
\end{equation}
As in the discrete case,  we define $R(f)$ as relative entropy setting
$g(p)$ to a continuous constant function.  In particular, the relative
entropy  of ${\cal P}(\phi)$  with respect  to a  uniform distribution
$\frac{1}{2\pi}$  \cite{sb06,sr07}   over  $\phi$  is   given  by  the
functional
\begin{equation}
R[{\cal P}(\phi)] = \int_0^{2\pi}d\phi~
{\cal P}(\phi)\log[2\pi {\cal P}(\phi)],
\label{eq:phient}
\end{equation}
where the $\log(\cdot)$ refers to the binary base.

We  define minimum  entropy states  with respect  to an  observable as
states that yield the minimum  entropy when the observable is measured
on  them.  In the  context of  relative entropy,  these states  can be
generalized  to what  may be  called maximum  knowledge  (MXK) states,
which are  applicable even when  the measured variable  is continuous.
For projector valued measurements  (PVMs), clearly any eigenstate is a
MXK state,  with a  corresponding entropy of  zero and knowledge  $R =
\log d$.  PVMs, projectors to  the eigenstates of a Hermitian operator
representing an observable, satisfy three axiomatic requirements: they
are positive  operators that form  a partition of unity;  further they
satisfy    the   orthonormalcy    condition    $\hat{P}_j\hat{P}_k   =
\delta_{jk}\hat{P}_j$,  where $\hat{P}_j$  is a  measurement operator.
The  last  property  implies  the  idempotency  of  projectors,  which
captures the idea that projective measurements are repeatable.  From a
quantum information perspective, it  is useful to consider generalized
measurements  in  which  the   operator  elements  $M_m$  may  not  be
orthonormal,   but   satisfy   the  completeness   condition   $\sum_m
M_m^{\dag}M_m  = I$  and $M^{\dag}_mM_m  \ge 0$  \cite{nc00}.   In the
context  of a  qubit,  for a  generalized  measurement, the  knowledge
corresponding  to a  MXK  state can  be  less than  1, i.e.,  $R(|{\rm
MXK}\rangle) \le 1$.   For a PVM, we have  $R(|{\rm MXK}\rangle) = 1$,
whereas a  POVM considered  here is a  measurement strategy  such that
$R(|{\rm  MXK}\rangle) < 1$.   The reason  is that  whereas PVM  is an
orthonormal  resolution  of  unity,  a POVM  forms  a  non-orthonormal
resolution  of  unity   \cite{holevo}. POVMs  are  useful
elsewhere, in  quantum information, as  general measurement strategies
for optimally distinguishing states \cite{nc00}. 

\begin{figure}
\begin{center}
\resizebox{0.75\columnwidth}{!}           {\includegraphics{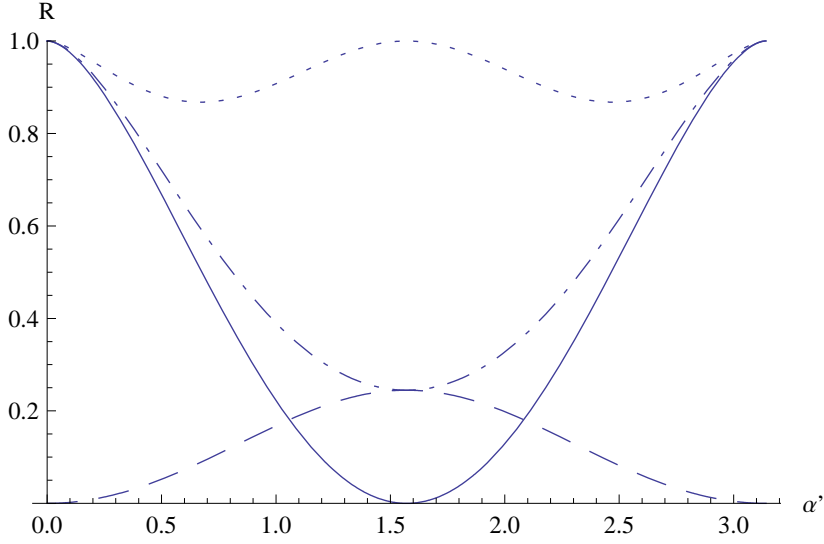}}
\caption{Plot of knowledge $R$  with respect to $\alpha^{\prime}$ of a
qubit      starting     in      an      atomic     coherent      state
$|\alpha^\prime,\beta^\prime\rangle$.  Knowledge $R$ is symmetric with
respect  to $\beta^\prime$,  which is  therefore not  depicted  in the
figure. The  individual curves are: $R_\phi  \equiv R[{\cal P}(\phi)]$
(dashed  curve),  $R_m  \equiv   R[{\cal  P}(m)]$  (bold  curve),  the
knowledge sum $R_\phi+R_m$ (dot-dashed) and the weighted knowledge sum
$R_S(\mu)  = \mu  R_\phi+R_m$  with $\mu=4.085$  (dotted curve).   The
maximum  of  $R_\phi$  is  not  1  but 0.245  bits,  which  occurs  at
$\alpha^{\prime}=\pi/2$,  i.e.,  the equatorial  states  on the  Bloch
sphere, which are thus the  MXK states or quasi-eigenstates of $\phi$.
The  minimum value of  $R_\phi= 0$  occurs at  $\alpha^{\prime}=0$ and
$\pi$, corresponding to the Wigner-Dicke states.}
\label{fig:minentcoh}
\end{center}
\end{figure}

A plot  of $R_\phi  \equiv R[{\cal P}(\phi)]$  for a  two-level atomic
system        in         an        atomic        coherent        state
$|\alpha^{\prime},\beta^{\prime}\rangle$  with  ${\cal  P}(\phi) =  {1
\over   2   \pi}\left[1   +   {\pi  \over   4}   \sin(\alpha^{\prime})
\cos(\beta^{\prime} - \phi)\right]$  \cite{sb06,sr07}, is given by the
dashed curve  in Figure (\ref{fig:minentcoh}).  We  note that $R_\phi$
has no dependence on $\beta^{\prime}$ because $\beta^\prime$ occurs in
${\cal  P}(\phi)$ only as  the translation  $\phi-\beta^{\prime}$, and
$R_\phi$   is  translation  invariant,   i.e.,  unchanged   under  the
transformation  $\phi  \longrightarrow \phi  +  \Delta$.  The  maximum
knowledge $R_\phi$  of about 0.245  occurs at $\alpha^{\prime}=\pi/2$.
The      corresponding      continuous      family      of      states
$|\pi/2,\beta^{\prime}\rangle$   forms   the   MXK  states   or   {\em
quasi-eigenstates}  of  the  phase  observable. These  are  equatorial
states     on     the     Bloch     sphere,    having     the     form
$\frac{1}{\sqrt{2}}(|0\rangle +  e^{i\phi_0}|1\rangle)$. That $R_\phi$
is less  that 1  for these  states reflects the  fact that  here phase
$\phi$ is a POVM.

In analogy  with the oscillator  case, the Wigner-Dicke  or excitation
states may be thought of  as `number states', thereby making $J_z$ the
`number   observable',  whose   distribution  is   $p(m)$,   given  in
Eq. (\ref{pnum}).  The `number' distribution given by
\begin{equation}
p(m) = \langle j,m|\rho^s(t)|j,m\rangle, \label{pnum} 
\end{equation}
is considered as complementary  to ${\cal P}(\phi)$ \cite{as96}. It is
of  interest to ask  whether they  are complementary  in the  sense of
MUBs.

In the manner of Eq. (\ref{eq:rf}), we can define $R_m \equiv R[p(m)]$
as  knowledge of the  number variable.  We note  that $J_z$  and phase
$\phi$ have a reciprocal behavior reminiscent of MUBs: the eigenstates
of $J_z$,  i.e., Wigner-Dicke states, correspond  to minimal knowledge
$R_\phi   (=  0)$,   as  seen   from  the   dashed  curve   in  Figure
(\ref{fig:minentcoh}).   This  can be  seen  by  noting  that for  the
Wigner-Dicke states $|j, \tilde{m} \rangle$, the phase distribution is
\cite{sb06}
\begin{equation}
{\cal P}(\phi) =  {2j+1 \over 2 \pi}  \left(\begin{array}{c}
2j \\ j + \tilde{m} \end{array} 
\right) {\cal B}\left[j + \tilde{m} + 1, j - \tilde{m} + 1 
\right] = \frac{1}{2\pi}, \label{pwd} 
\end{equation}
where ${\cal B}$  stands for the Beta function.   Thus, it follows via
Eq.    (\ref{eq:phient})  that   the   knowledge  $R_\phi$   vanishes.
Conversely, we  note that the  states which minimize $R_\phi$  are the
Wigner-Dicke states. To see this,  we observe that if ${\cal P}(\phi)$
is constant,  then in  Eq. (\ref{2a.7}), each  term in  the summation,
which  is  proportional  to  $e^{i(m-n)\phi}$,  must  individually  be
independent  of $\phi$. Since  $\phi$ is  arbitrary, this  is possible
only  if  $m=n$,   i.e.,  the  state  $\rho^s$  is   diagonal  in  the
Wigner-Dicke basis.   Thus, MXK states of $m$  correspond precisely to
minimum knowledge (MNK) states of $\phi$.

The plot of  relative entropy $R_m$ for all  atomic coherent states is
given  by  the  bold   curve  in  Figure  (\ref{fig:minentcoh}).   The
equatorial  states  on  the  Bloch  sphere, the  MXK  of  $\phi$,  are
precisely  equivalent  to the  MNK  states  of  $m$ (characterized  by
$R_m=0$), as can be seen from  comparing the dashed and bold curves in
Figure (\ref{fig:minentcoh}).   Thus number and phase  share with MUBs
the  reciprocal property  that maximum  knowledge  of one  of them  is
simultaneous  with minimal knowledge  of the  other, but  differs from
MUBs in  that the  maximum possible knowledge  of $\phi$ is  less than
$\log(d) = 1$ bit, essentially on account of its POVM nature.

Two variables form a quasi-MUB if  any MXK state of either variable is
an MNK state of the other, where the knowledge of the MXK state may be
less than $\log d$ bits.   Thus, $J_z$ and $\phi$ are quasi-MUB's (but
not MUB's), and are complementary in the extended sense.

From  the dot-dashed  curve in  this  Figure, we  numerically find  an
expression of the uncertainty principle to be
\begin{equation}
\label{eq:1bit}
R_T \equiv R_\phi + R_m \le 1
\end{equation}
for all  states (pure or in general mixed) 
    in      $\mathbb{C}^2$,      in
analogy with    Eq.   (\ref{eq:ra}).    The
inequality is saturated for the Wigner-Dicke states.

As  an   expression  of   the  uncertainty  principle,   the  relation
(\ref{eq:1bit}) still leaves some  room for improvement.  First, it is
not a  tight bound.  In  particular, for equatorial states  it permits
$R_\phi$ to be as high as 1,  whereas as seen from the dashed curve in
Figure  \ref{fig:minentcoh},  the  maximum  value of  $R_\phi  \approx
0.245$.   We  note  that  the  bound cannot  be  tightened  simply  by
decreasing the  r.h.s, since it is saturated  for Wigner-Dicke states.
Further, the  variable $\phi$ takes values in  the interval $[0,2\pi]$
irrespective of  the dimensionality of the Hilbert  space, unlike $m$,
which takes $d$ values.   Consequently, $R_\phi$, unlike $R_m$, is not
seen  to  be bounded  by  the  dimension of  the  Hilbert  space in  a
straightforward  way.  To  see  that in  general  $R[p(x)]$  increases
without bound, consider the probability distribution $p(x)=x_0 > 1$ in
$x \in  [0,\frac{1}{x_0}]$ and $p(x)=0$ in  $x \in (\frac{1}{x_0},1]$,
for which we find $R(p(x)||1) =\log x_0$.

One way to address these  problems is to generalize (\ref{eq:1bit}) to
a family of  inequalities, parametrized by
$\mu>0$, of the form
\begin{equation}
\label{eq:2bit}
R_S(\mu) \equiv \mu R_\phi + R_m \le 1
\end{equation}
for all states in $\mathbb{C}^2$.  We find that the largest value
of  $\mu$ such that  inequality (\ref{eq:2bit})  is satified  over all
state space  is $(r_\phi)^{-1} \approx  4.085$, where $r_\phi$  is the
value of $R_\phi$ for the equatorial states, the MXK states of $\phi$.
A  plot of $R_S(1/r_\phi)$  over pure  states is  shown as  the dotted
curve  in   Figure  \ref{fig:minentcoh}.   Comparing   this  with  the
dot-dashed   curve  in  Figure   \ref{fig:minentcoh},  we   find  that
$R_S(\mu)$ is bounded more tightly than $R_T \equiv R_S(1)$.

From Figure  \ref{fig:minentcoh}, we  find that  the two
Wigner-Dicke states and all equatorial  states may be regarded as {\em
coherent} with respect to the number-phase pair, in that they maximize
the knowledge sum  and are thus closest to  classical states.  We note
of course  that this  definition of state  coherence differs  from the
conventional one  for atomic states,  defined with respect  to angular
momentum operators. Unless  we use $\mu R_\phi$ in  place of $R_\phi$,
only  the Wigner-Dicke  states could  be  called coherent  in the  new
sense. 

We now  briefly extend  the entropic version  of complementarity  to a
higher  spin  system, which  is  seen to  present  a  new feature.  We
consider a spin-3/2 (four-level)  system, whose general state is given
by the ansatz
\begin{equation}
|\psi\rangle = r_\alpha e^{i\theta_\alpha}|\frac{3}{2},-\frac{3}{2}\rangle +
r_\beta e^{i\theta_\beta}|\frac{3}{2},-\frac{1}{2}\rangle +
r_\gamma e^{i\theta_\gamma}|\frac{3}{2},+\frac{1}{2}\rangle +
r_\delta |\frac{3}{2},+\frac{3}{2}\rangle)
\label{eq:ansatz}
\end{equation}
where $r_\alpha^2 + r_\beta^2+r_\gamma^2+r_\delta^2=1$, and a global
phase is omitted. Using Eq. (\ref{eq:ansatz}) in Eq. (\ref{2a.7}),
we find
\begin{eqnarray}
P(\phi) &=& \frac{1}{\pi}\left[\frac{1}{2} + 
\left(\frac{15\pi r_\alpha r_\beta}{32\sqrt{3}}\right)
\cos(\phi-\theta_\alpha\theta_\beta) +
\left(\frac{r_\alpha r_\gamma}{\sqrt{3}}\right)
\cos(2\phi-\theta_\alpha \theta_\gamma) \right. \nonumber \\ &+& 
\left(\frac{ 3\pi r_\alpha r_\delta}{32}\right)
\cos(3\phi-\theta_\alpha) +
\left(\frac{ 9\pi r_\beta r_\gamma}{32}\right)
\cos(\phi-\theta_\beta+\theta_\gamma) +
\left(\frac{r_\beta r_\delta}{\sqrt{3}}\right)
\cos(2\phi-\theta_\beta) \nonumber \\ &+&
\left. \left(\frac{15\pi r_\gamma r_\delta}{32\sqrt{3}}\right)
\cos(\phi-\theta_\gamma)\right].
\end{eqnarray}

As       before,       we       compute       `number'       knowledge
$R_m(r_\alpha,r_\beta,r_\gamma)$  using Eq.  (\ref{eq:rf}),  and phase
knowledge              $R_\phi(r_\alpha,r_\beta,             r_\gamma,
\theta_\alpha,\theta_\beta,\theta_\gamma)$                        using
Eq. (\ref{eq:phient}).   It may be  verified that for  `number' states
(for which $r_\alpha$ or $r_\beta$  or $r_\gamma$ or $r_\delta$ is 1),
$R_\phi=0$.   In  fact,  it  may  be  seen  from  Eqs.   (\ref{2a.6}),
(\ref{2a.7})  and  (\ref{pwd}),  that  a general  property  of  atomic
systems  is that a  Wigner-Dicke state  is equivalent  to a  MNK phase
state  in  any  finite  dimension.   On the  other  hand,  numerically
searching over  all possible states of the  form (\ref{eq:ansatz}), we
find that the maximum value  0.86 bits of $R_\phi$ occurs
at                      $\Psi(r_\alpha=0.36,r_\beta=0.61,r_\gamma=0.61,
\theta_\alpha=\pi,\theta_\beta=0,\theta_\gamma=\pi)$, which  is not an
equal amplitude superposition of `number' states.  Thus,
remarkably, for the spin-3/2 case,  MXK phase states do not correspond
to MNK `number'  states, even though the converse  is true.  We expect
that  this unidirectionally  (as against  mutually)  unbiased behavior
will  persist  even  for  higher  spin systems.   Phase  and  `number'
therefore do not here form a quasi-MUB as defined for the single qubit
case, and  may be considered complementary  only in an  even more weak
sense.   This is in  contrast to  the case  where the  observables are
Hermitian,  where five  MUBs are  known  to exist  in four  dimensions
\cite{dur05}.

As in the two-level case, one way to address this  problem is 
to generalize (\ref{eq:1bit}) to
a  family of  inequalities, parametrized by
$\mu_2>0$, of the form
\begin{equation}
\label{eq:4bit}
R_S(\mu_2) \equiv \mu_2 R_\phi + R_m \le 2
\end{equation}
over  all states in  $\mathbb{C}^4$.  Our  strategy is  to numerically
search over all states of  the form (\ref{eq:ansatz})-- other than the
Wiger-Dicke  states,  where  $R_m=2$  and  $R_\phi=0$--  in  order  to
determine  the   largest  value   of  $\mu_2$  such   that  inequality
(\ref{eq:4bit}) is  {\em just} satified, i.e., the  inequality must be
satisfied at  all points, with the  equality being valid  for at least
one point.  By  this method, we find $\mu_2 =  1.973$ with the maximum
$R_S(\mu_2)$     of      2     occuring     at      $\psi_p     \equiv
\psi(r_\alpha=0.24,r_\beta=0.64,r_\gamma=0.68,
\theta_\alpha=\pi,\theta_\beta=0,\theta_\gamma=\pi)$.   As states that
maximize the  knowledge sum $R_S(\mu_2)$,  we may regard  $\psi_p$ and
the  Wigner-Dicke states  as  coherent states  from  the viewpoint  of
number-phase entropy.  

\section{Application to open systems \label{sec:open}}

Here  we study the  effect of  noise coming  from open  quantum system
effects, on  the atomic number-phase complementarity  developed in the
previous   section.   The   noise  effects   we  consider   come  from
non-dissipative  as well  as  dissipative interactions  of the  atomic
system  $S$  with its  environment  which is  modelled  as  a bath  of
harmonic   oscillators   starting   in   a  squeezed   thermal   state
\cite{bg07,gp,sqgen}. This  enables us to consider the  effect of bath
squeezing  on the complementarity.   We briefly  recapitulate previous
work \cite{sr07,bg07,sb06,gp,sqgen}  related to the  effect of various
noisy channels  on the `number'  and phase distributions.   In Section
\ref{sec:openphase}  we  consider  the  effect of  the  phase  damping
channel   which  is   the  information   theoretic  analogue   of  the
non-dissipative  open system  effect \cite{bg07,gp}  while  in Section
\ref{sec:opengen} we  consider the effect of  the squeezed generalized
amplitude damping channel which  is the information theoretic analogue
of the  dissipative open system  effect \cite{gp,sqgen}.  Intuitively,
one would  expect that open system effects,  like measurements, cannot
increase the  knowledge sum. Interestingly,  we find that this  is not
true for certain regimes of the squeezed generalized amplitude damping
channel.

\subsection{Phase damping channel \label{sec:openphase}}
The  `number' and  phase distributions  for a  qubit starting  from an
atomic   coherent   state  $|\alpha^\prime,\beta^\prime\rangle$,   and
subjected to  a phase  damping channel due  to its interaction  with a
squeezed thermal bath, are \cite{sr07,bg07,sb06}
\begin{eqnarray}
p(m) &=& \left( \begin{array}{c} 2j \\ j+m\end{array}\right) 
(\sin(\alpha^{\prime}/2))^{2(j+m)}
(\cos(\alpha^{\prime}/2))^{2(j-m)} \nonumber \\
{\cal P}(\phi) &=& {1 \over 2 \pi}\left[1 + {\pi \over 4} 
\sin(\alpha^{\prime}) \cos(\beta^{\prime} + \omega t - \phi) e^{- 
(\hbar \omega)^2 \gamma(t)}\right]. \label{2a.9} 
\label{eq:atomcohqnd}
\end{eqnarray}
$R_\phi$ (Eq. (\ref{eq:phient})) is invariant under the translation of
$\phi \longrightarrow \phi+a$. Setting  $a= -\beta^\prime - \omega t$,
we find  that $R_\phi$ is independent of  $\beta^\prime$. A derivation
of Eq.  (\ref{eq:atomcohqnd}) can  be found in Refs. \cite{sb06,sr07}.
For completeness,  the expression for $\gamma(t)$  in Eq. (\ref{2a.9})
is  given in  Appendix \ref{secap:qnd}  and  the physical
process underlying the phase damping channel discussed. 

\begin{figure}
\begin{center}
\resizebox{0.75\columnwidth}{!}
{\includegraphics{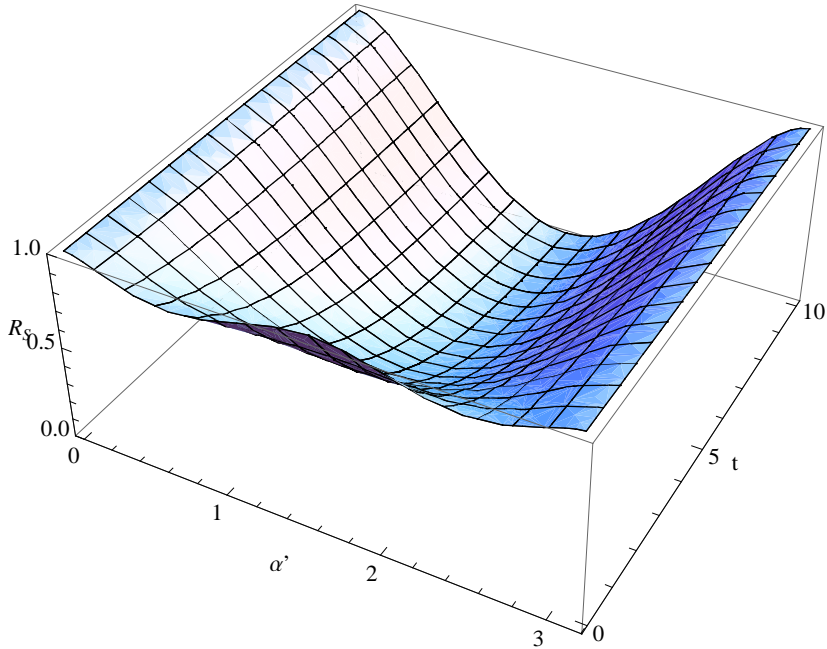}}
\caption{Plot of $R_S  = \mu R_\phi+R_m$ for a  qubit starting from an
atomic  coherent  state  $|\alpha^{\prime},\beta^\prime=\pi/4\rangle$,
subjected to a phase damping channel, with temperature (in units where
$\hbar\equiv  k_B \equiv  1$) $T=2$,  $\gamma_0=0.025$, $\omega_c=100$
and  $\omega=1.0$,  with  respect  to  bath  exposure  time  and  bath
squeezing parameters  (Eq. (\ref{eq:a})) $r= 1$ and  $a=0.0$.  For the
phase  damping  channel,  $R_m$  remains invariant.   Both  $R_m$  and
$R_\phi$, and hence $R_S$, are independent of $\beta^{\prime}$.}
\label{fig:sqminfiqnd}
\end{center}
\end{figure}

Figure \ref{fig:sqminfiqnd} depicts the  effect of phase damping noise
on  the knowledge  sum $R_S$.   Comparing it  with the  noiseless case
(dotted curve  in Figure \ref{fig:minentcoh}), we find  a reduction in
the  total  knowledge  $R_S$,   as  expected.   It  follows  from  Eq.
(\ref{eq:atomcohqnd}) that  $R_m$ remains unaffected  under the action
of this channel.   Thus, the effect of noise on  $R_S$ is due entirely
to its  effect on $R_\phi$, which  decreases in the  presence of noise
{\em for  all pure  states} (because  $\beta^\prime$ does
not play any role and  because the plot represents all possible values
of $\alpha^\prime$).

Figure \ref{fig:sqminfiqnd} shows
that squeezing has the detrimental effect of impairing phase knowledge
for all regimes of the parameter space.  This is in marked contrast to
the  case  of  the   squeezed  generalized  amplitude  damping  noise,
discussed  in  Section  (\ref{sec:opengen}).   Thus,  squeezing,  like
temperature,   has  the  overall   detrimental  effect   of  impairing
$R_S$. This  is consistent  for the case  of a QND  interaction (which
generates a  phase damping channel \cite{bg07,gp}) of  the system with
its environment, i.e.,  $[H_S, H_{SR}] = 0$, as  also corrorborated by
the  observation that  squeezing and  temperature  concurrently impair
geometric  phase \cite{gp}  and phase  diffusion  \cite{sb06,sr07} and
suggests  that squeezing,  like temperature,  should  adversely affect
channel capacity for phase damping noise.

\subsection{Squeezed generalized amplitude damping channel 
\label{sec:opengen}}
The  `number' and  phase distributions  for a  qubit starting  from an
atomic   coherent   state  $|\alpha^\prime,\beta^\prime\rangle$,   and
subjected  to   a  squeezed  generalized   amplitude  damping  channel
\cite{sqgen} due to its interaction  with a squeezed thermal bath, are
\cite{sr07},
\begin{equation}
p(m=1/2,t) 
= \frac{1}{2}\left[\left(1-\frac{\gamma_0}{\gamma^\beta}\right)
+ \left(1+\frac{\gamma_0}{\gamma^\beta}\right)e^{-\gamma^\beta t}\right]
\sin^2(\alpha^\prime/2) +
\frac{\gamma_-}{\gamma^\beta}\left(1-e^{-\gamma^\beta t}\right)
\cos^2(\alpha^\prime/2), 
\label{eq:pmc}
\end{equation}
and
\begin{eqnarray}
{\cal  P}(\phi) &=&  \frac{1}{2  \pi} \left[1  + \frac{\pi}{4  \alpha}
\sin(\alpha^{\prime})   \Big\{\alpha    \cosh(\alpha   t)\cos(\phi   -
\beta^{\prime}) +  \omega \sinh(\alpha t)  \sin(\phi - \beta^{\prime})
\right. \nonumber\\ & & \left.  - \gamma_0 \chi \sinh(\alpha t) \cos(\Phi
+  \beta^{\prime}  +   \phi)  \Big\}  e^{-\frac{\gamma^{\beta}  t}{2}}
\right].  \label{3p}
\end{eqnarray}
A derivation  of Eqs.  (\ref{eq:pmc})  and (\ref{3p}) can be  found in
Refs.   \cite{sr07}.  For  completeness, the  parameters  appearing in
these  equations are given  in Appendix  \ref{secap:disi} 
where a brief  discussion of the physical process  behind the squeezed
generalized amplitude damping channel is also made. 

\begin{figure}
\begin{center}
\resizebox{0.98\columnwidth}{!}
{\includegraphics{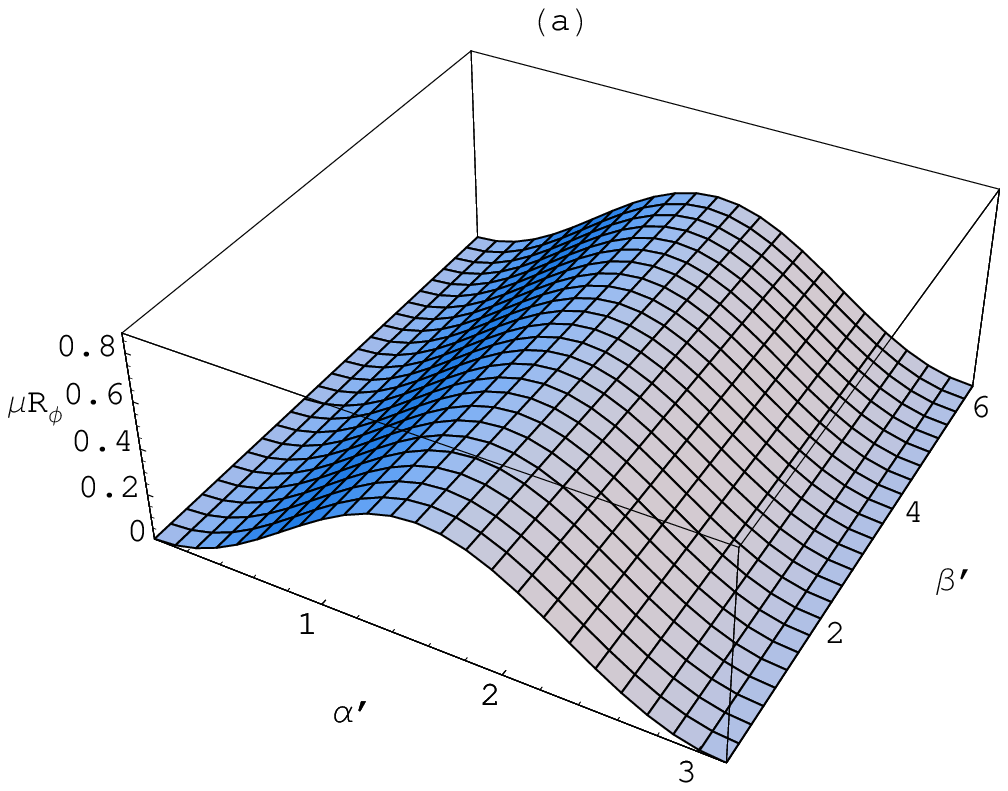}
\hfill
\includegraphics{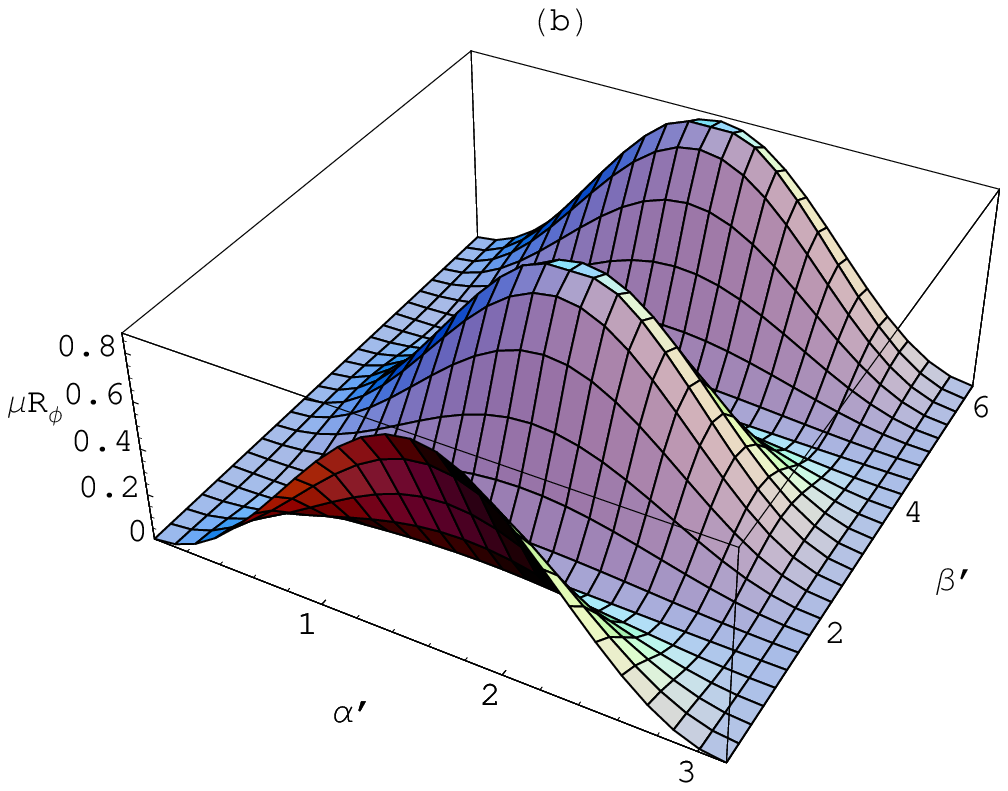}}
\caption{Plot of $\mu  R_\phi$ (scaled $R_\phi$ [Eq. (\ref{eq:2bit})])
for    a   qubit    starting   from    an   atomic    coherent   state
$|\alpha^{\prime},\beta^{\prime}\rangle$, and  subjected to a squeezed
generalized amplitude damping channel with temperature (in units where
$\hbar\equiv  k_B \equiv 1$)  $T=300$, $\gamma_0=0.01$,  bath exposure
time $t=0.1$,  and $\omega=1.0$.  Figure (a)  is for the  case of zero
bath  squeezing, and  (b) for  the case  of bath  squeezing parameters
(Eq. (\ref{eq:M})) $r= 1$  and $\Phi=\pi/8$.  Squeezing has the effect
of  breaking translation  symmetry in  $\beta^{\prime}$  and improving
phase knowledge (i.e., reducing phase uncertainty) for certain regimes
of the parameter space.}
\label{fig:sqminfi}
\end{center}
\end{figure}

Figures  \ref{fig:sqminfi}(a) and  (b) depict  the effect  of squeezed
generalized    amplitude    damping     noise    on    $\mu    R_\phi$
(Eq. (\ref{eq:2bit})), without  and with bath squeezing, respectively.
Comparing them  with the noiseless case  of Figure \ref{fig:minentcoh}
(which, it  may be noted, is  unscaled by $\mu$), we  find as expected
that  noise  impairs   phase  knowledge.   However,  comparing  Figure
\ref{fig:sqminfi}(b)  with  (a),  we   find  that  squeezing  has  the
beneficial effect of relatively  improving phase knowledge for certain
regimes  of  the  parameter  space,  and  the  detrimental  effect  of
relatively impairing  them in others.   This property can be  shown to
improve  the classical channel  capacity \cite{sqgen}.   Further, bath
squeezing is  seen to  render $R_\phi$ dependent  on $\beta^{\prime}$,
because,  as evident  from Eq.   (\ref{3p}), $\beta^\prime$  no longer
appears as a translation in $\phi$ when the squeezing parameter $\chi$
(Eq.  (\ref{eq:M}))  is non-vanishing.  On the other  hand, it follows
from Eq.  (\ref{eq:pmc}) that  $R_m$ is independent of $\beta^\prime$,
so that $R_S$ is dependent on $\beta^\prime$.  This stands in contrast
to  that of  the phase  damping channel,  where inspite  of squeezing,
$R_S$  remains  independent   of  $\beta^{\prime}$  and,  furthermore,
squeezing impairs knowledge of $\phi$  in all regimes of the parameter
space.

\begin{figure}
\begin{center}
\resizebox{0.98\columnwidth}{!}
{\includegraphics{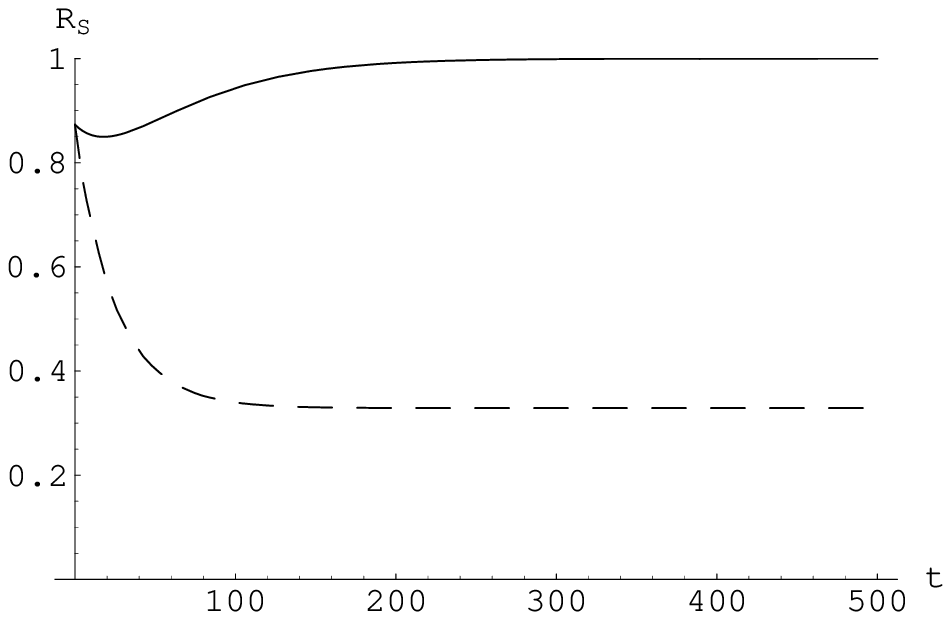}
\hfill
\includegraphics{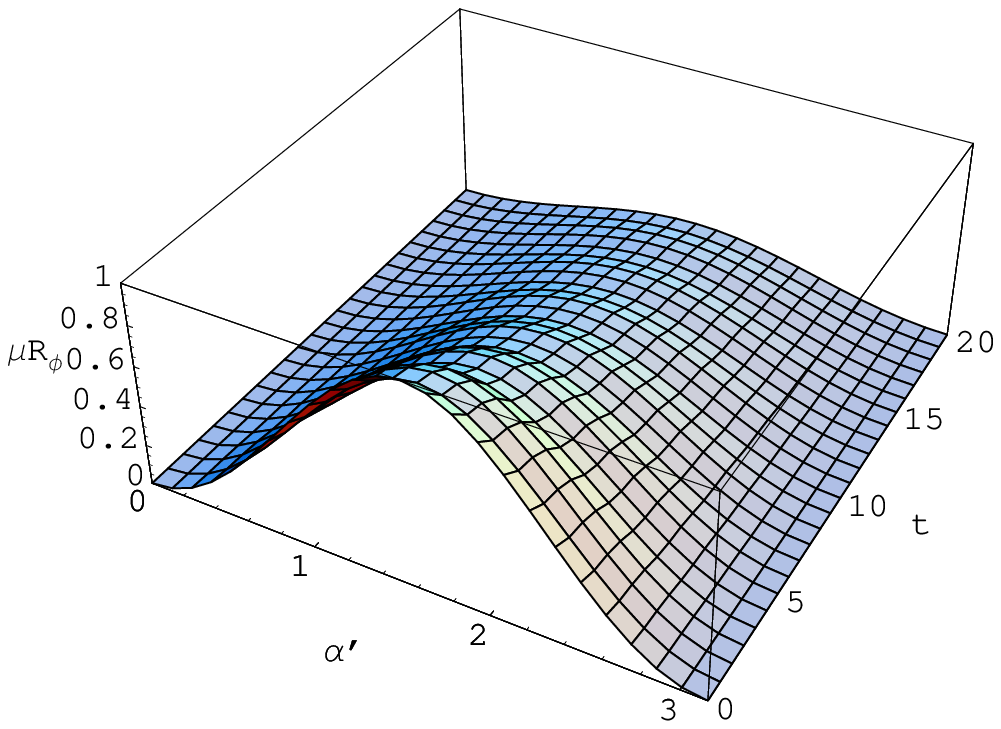}}
\caption{(a) Plot of $R_S \equiv  \mu R_\phi+R_m$ with respect to time
  for   a    qubit   starting   from   an    atomic   coherent   state
  $|\alpha^{\prime}=\pi/4,\beta^{\prime}=\pi/4\rangle$  and  subjected
  to   a  squeezed   generalized  amplitude   damping   channel.   The
  temperature  (in  units where  $\hbar\equiv  k_B  \equiv 1$)  $T=0$,
  $\gamma_0=0.025$, and  $\omega=1.0$.  The bath  squeezing parameters
  (Eq.  (\ref{eq:M})) are $\Phi=0$, and $r = 0$ ($r=0.5$) for the case
  of  bold  (dashed)  curves.   (b)  Plot  of  $\mu  R_\phi$  for  all
  $\alpha^\prime$ as  a function of time  for a qubit  starting from a
  coherent   state  $|\alpha^\prime,\pi/2\rangle$   and   subjected  a
  squeezed  amplitude  damping channel  with  the  same parameters  as
  above,   with    $T=\Phi=0$,   $r=0.5$   and    $\gamma=0.05$.   The
  $R_\phi$-decreasing effect of increasing $T$ or $r$ is qualitatively
  the same.}
\label{fig:usq}
\end{center}
\end{figure}

A  point worth noting is  that, in contrast  to the phase
damping channel, in a  squeezed generalized amplitude damping channel,
$R_m$ and $R_S$ are  not necessarily non-increasing functions of time.
Figure  \ref{fig:usq}(a) depicts the  effect of squeezed
generalized amplitude  damping channel on  $R_S$, by bringing  out the
behavior of  $R_S$ as  a function of  bath exposure time.   The dashed
curve shows that  squeezing has a detrimental effect  on the knowledge
sum, as  one would  usually expect. A  surprising departure  from this
behavior  may  be  noted  for  the  case  of  the  bold  curve,  which
corresponds  to  the  action  of  a dissipative  interaction  with  an
unsqueezed vacuum bath, where the  knowledge sum $R_S$ increases to 1.
This counterintuitive behavior is  due to the quantum deleting action,
a contractive map whereby any  initial state, including a mixed state,
is  asymptotically prepared in  the pure  state $|\half,-\half\rangle$
for  vanishing temperature, and a  mixture of $|0\rangle$
and $|1\rangle$  states for  finite temperature, where  the asymptotic
mixture is  determined purely by  the environmental parameters  of $T$
and  $r$, and  not  by  the system's  initial  state \cite{qdele}.   A
similar effect was  noted in \cite{gp93}, where in  a study of quantum
state diffusion  of an open  system it was  shown that for  a specific
noise, due to a  particular system-reservoir interaction, there can be
a reduction in the quantum dispersion entropy leading to localization.

It  follows from  the complementarity  relation  Eq.  (\ref{eq:2bit}),
that in the asymptotic limit  of the deleting action, $R_\phi$ goes to
0  for  both $|0\rangle$  and  $|1\rangle$,  and  hence also,  by  the
convexity of $R$, for any mixture that is diagonal in this basis. More
generally,  it  is seen  from  Figure  \ref{fig:usq}(b)  that for  all
initial  pure states,  $R_\phi$ falls  monotonically.  This  is  to be
expected since  this noise prepares  an asymptotic state that  lies on
the  $z$-axis of  the Bloch  sphere,  which implies  by the  convexity
property of  $R_\phi$ and the fact  that $R_\phi=0$ for  the north and
south pole states, that asymptotically $R_\phi=0$ for {\em all initial
  pure states}.  

\section{Conclusions \label{sec:concl}}
In this work, we have investigated the number-phase complementarity in
atomic  systems from an  entropic perspective  through the  number and
phase   distributions.   Here  number   distribution  refers   to  the
probability distribution of measurement in the Wigner-Dicke basis (Eq.
(\ref{pnum})),    while    phase    distribution   is    defined    by
Eq. (\ref{2a.7}).  We derive an  uncertainty principle in terms of the
Kullb\"ack-Leibler or  relative entropy $R$  of number and  phase with
respect  to a uniform  distribution. Since  $R$ can  be regarded  as a
measure of  knowledge of a  random variable, the  entropic uncertainty
principle  takes the  form of  an  upper bound  on the  sum of  number
knowledge  ($R_m$)  and phase  knowledge  ($R_\phi$).   The choice  of
relative entropy over  Shannon entropy was motivated by  the fact that
the  latter  is  not  strictly  positive when  applied  to  continuous
probability distributions.

In the single-qubit case, number  and phase are regarded as quasi-MUBs
in  the sense  that any  state  maximizing knowledge  of one  variable
simultaneously  minimizes knowledge  of the  other, but  maximum phase
knowledge is strictly less than 1  bit (and less than $\log d$ bits in
$d$ dimensions).

Since  $R_\phi$  is strictly  less  than  one,  the relative  entropic
formulation of  the uncertainty  principle does not  tightly constrain
$R_m$.  We define a family of inequalities, parametrized by $\mu$ (Eq.
(\ref{eq:2bit})),  that  improves  the  upper bound  on  $R_m$.   When
$\mu=1$, we obtain Eq. (\ref{eq:1bit}), and get the tightest bound for
equatorial  states (with  the  right hand  side  saturated) when  $\mu
\approx 4.085$. We  briefly study the extension of  the above concepts
to a four-level  system, where we find that the  sense in which number
and phase  are said  to be complementary  must be further  weakened to
include unidirectional (but  not mutual) unbiasedness.  In particular,
whereas phase is unbiased with  respect to number, the converse is not
true.

Finally, we study the complementary  behavior of number and phase of a
qubit  subjected to  the influence  of its  environment.  For  a qubit
starting        from       an       atomic        coherent       state
$|\alpha^\prime,\beta^\prime\rangle$,  the   translation  symmetry  of
$R_\phi$ in $\beta^\prime$ is  broken by the introduction of squeezing
in the  bath, for  the case of  a dissipative  system-bath interaction
(Figure   \ref{fig:sqminfi}(b)),   but   not   in  the   case   of   a
non-dissipative  interaction.   In the  case  of  a purely  decohering
interaction, characterized  by a phase  damping channel, we  find that
noise  invariably impairs  the knowledge  sum for  these complementary
variables (Eq.   (\ref{eq:2bit})), as expected.  However,  in the case
of  a squeezed generalized  amplitude channel,  the knowledge  sum can
increase in certain regimes.  As a particularly dramatic illustration,
when       an        initially       maximally       mixed       state
$\half(|\half,\half\rangle\langle\half,\half|                         +
|\half,{\rm-}\half\rangle\langle\half,{\rm-}\half|)$  is  subjected to
an unsqueezed vacuum bath, $R_S$ rises from 0 to 1 asymptotically.

These results could be  potentially useful for applications in quantum
communication and  quantum cryptography \cite{gis02}  involving atomic
systems.  The  present work  brings forth a  number of  open questions
concerning an information theoretic study of complementarity in atomic
systems involving continuous-valued POVMs, of which we list some here.
Of immediate interest  is the question whether the  Shannon entropy of
${\cal  P}(\phi)$ remains  positive for  all possible  pure  and mixed
states. If yes, then one may revert back from the use of the knowledge
variable  $R$  to  that  of  entropy  $S$.  Also  of  interest  is  to
analytically derive the bounds on the weighted knowledge sum, which we
have obtained here numerically.  Finally, it is of interest to explore
the  full  scope  and  implications  of one-way  biasedness,  and  its
connection to complementarity.

\appendix

\section{Phase damping channel  \label{secap:qnd}}

Consider the Hamiltonian
\begin{eqnarray}
H & = & H_S + H_R + H_{SR} \nonumber \\ & = & H_S + 
\sum\limits_k \hbar \omega_k b^{\dagger}_k b_k + H_S 
\sum\limits_k g_k (b_k+b^{\dagger}_k) + H^2_S \sum\limits_k 
{g^2_k \over \hbar \omega_k}. 
\end{eqnarray} 
Here  $H_S$, $H_R$  and $H_{SR}$  stand  for the  Hamiltonians of  the
system,      reservoir      and     system-reservoir      interaction,
respectively.  $H_S$ is  a  generic system  Hamiltonian  which can  be
specified depending on the physical situation.  $b^{\dagger}_k$, $b_k$
denote  the  creation and  annihilation  operators  for the  reservoir
oscillator  of frequency  $\omega_k$,  $g_k$ stands  for the  coupling
constant (assumed  real) for the  interaction of the  oscillator field
with the system. The last term on  the right-hand side of Eq. (1) is a
renormalization inducing `counter  term'. Since $[H_S, H_{SR}]=0$, the
Hamiltonian (1) is of QND type. The system-plus-reservoir composite is
closed and hence obeys a unitary evolution given by
\begin{equation}
\rho (t) = e^{- iHt / \hbar} \rho (0) e^{iHt / \hbar} , 
\end{equation}
where
\begin{equation}
\rho (0) = \rho^s (0) \rho_R (0),
\end{equation}
i.e., we  assume separable initial  conditions.  Here $\rho_R  (0)$ is
the initial  density matrix  of the  reservoir which we  take to  be a
squeezed thermal bath given by
\begin{equation}
\rho_R(0) = S(r, \Phi) \rho_{th} S^{\dagger} (r, \Phi), \label{rhorin}
\end{equation}
where
\begin{equation}
\rho_{th} = \prod_k \left[ 1 - e^{- \beta \hbar \omega_k} 
\right] e^{-\beta \hbar \omega_k b^{\dagger}_k b_k} \label{rhoth}
\end{equation}
is the  density matrix  of the thermal  bath at temperature  $T$, with
$\beta \equiv 1/(k_B T)$, $k_B$ being the Boltzmann constant, and
\begin{equation}
S(r_k, \Phi_k) = \exp \left[ r_k \left( {b^2_k \over 2} e^{-2i 
\Phi_k} - {b^{\dagger 2}_k \over 2} e^{2i \Phi_k} \right) 
\right] \label{sqop}
\end{equation}
is  the squeezing operator  with $r_k$,  $\Phi_k$ being  the squeezing
parameters \cite{cs85}.   Here we  take the system  to be  a two-level
atomic system, with the Hamiltonian
\begin{equation}
H_S = {\hbar \omega \over 2} \sigma_z, \label{4a1}
\end{equation}
$\sigma_z$ being  the usual Pauli matrix.  The  reduced density matrix
of  the  system,  in the  basis  of  the  Wigner-Dicke states  $|j,  m
\rangle$, after time $t$ is \cite{gp}
\begin{equation}
\rho^s_{m,n}(t) = 
\pmatrix {\cos^2({\theta_0 \over 2})
& {1 \over 2} \sin(\theta_0)
e^{-i (\omega t + \phi_0)} e^{-(\hbar \omega)^2 \gamma(t)}
\cr {1 \over 2} \sin(\theta_0) 
e^{i(\omega t + \phi_0)} e^{-(\hbar \omega)^2 \gamma(t)}
& \sin^2({\theta_0 \over 2})}, \label{4a4}
\end{equation}
from which the Bloch vectors can be extracted to yield
\begin{eqnarray}
\langle \sigma_x (t) \rangle &=& \sin(\theta_0) \cos(\omega t + 
\phi_0) e^{-(\hbar \omega)^2 \gamma(t)}, \nonumber\\ \langle 
\sigma_y (t) \rangle &=& \sin(\theta_0) \sin(\omega t + \phi_0) 
e^{-(\hbar \omega)^2 \gamma(t)}, \nonumber\\ \langle \sigma_z 
(t) \rangle &=& \cos(\theta_0). \label{4a5} 
\end{eqnarray}
Here $\gamma(t)$ comes due to the interaction with the environment and
for the case of an Ohmic bath with spectral density
\begin{equation}
I(\omega) = {\gamma_0 \over \pi} \omega e^{-\omega/\omega_c}, 
\label{2.5} 
\end{equation}
where $\gamma_0$ and $\omega_c$ are two bath parameters characterizing
the quantum  noise, it can shown  that using Eq.   (\ref{2.5}) one can
obtain \cite{bg07} in the $T = 0$ limit,
\begin{equation}
\gamma (t)  =  {\gamma_0 \over 2\pi} \cosh (2r) \ln 
(1+\omega^2_c t^2) - {\gamma_0 \over 4\pi} \sinh (2r) \ln 
\left[ {\left( 1+4\omega^2_c(t-a)^2\right) \over \left( 1+ 
\omega^2_c (t-2a)^2 \right)^2} \right]  - 
{\gamma_0 \over 4\pi} \sinh (2r) \ln (1+4a^2\omega^2_c) , 
\label{2.7} 
\end{equation}
where the resulting  integrals are defined only for $t  > 2a$.  In the
high $T$ limit, $\gamma (t)$ can be shown to be \cite{bg07}
\begin{eqnarray} 
\gamma (t) & = & {\gamma_0 k_BT \over \pi \hbar \omega_c} \cosh 
(2r) \left[ 2\omega_c t \tan^{-1} (\omega_c t) + \ln \left( {1 
\over 1+\omega^2_c t^2} \right) \right] - 
{\gamma_0 k_BT \over 2\pi \hbar \omega_c} \sinh (2r) \nonumber \\
&\times& \Bigg[ 
4\omega_c (t-a) \tan^{-1} \left( 2\omega_c (t-a) \right) 
- 4\omega_c (t-2a) \tan^{-1} \left( \omega_c 
(t-2a) \right) + 4a\omega_c \tan^{-1} \left( 2a\omega_c \right) \nonumber \\
&+& \ln \left( {\left[ 1+\omega^2_c (t-2a)^2 
\right]^2 \over \left[ 1+4\omega^2_c (t-a)^2 \right]} \right) + 
\ln \left( {1 \over 1+4a^2\omega^2_c} \right) \Bigg] , 
\label{eq:gamma} 
\end{eqnarray} 
where, again, the  resulting integrals are defined for  $t > 2a$. Here
we have for simplicity taken the squeezed bath parameters as
\begin{eqnarray} 
\cosh \left( 2r(\omega) \right) & = & \cosh (2r),~~ \sinh 
\left( 2r (\omega) \right) = \sinh (2r), \nonumber\\ \Phi 
(\omega) & = & a\omega, \label{eq:a} 
\end{eqnarray} 
where $a$ is a constant depending upon the squeezed bath.  The results
pertaining to a thermal bath  can be obtained from the above equations
by  setting   the  squeezing  parameters  $r$  and   $\Phi$  to  zero.
$\sigma_x$, $\sigma_y$, $\sigma_z$ are the standard Pauli matrices. It
can be easily seen from the  above Bloch vector equations that the QND
evolution  causes a  coplanar, fixed  by the  polar  angle $\theta_0$,
in-spiral  towards the  $z$-axis of  the  Bloch sphere.   This is  the
characteristic of a phase-damping channel
\cite{nc00}. \\ \ \\

\section{Squeezed generalized amplitude damping channel
  \label{secap:disi}}  

Here the reduced  dynamics of the two level  atomic system (\ref{4a1})
interacting with a squeezed thermal  bath under a weak Born-Markov and
rotating wave  approximation is studied.   This implies that  here the
system interacts with its environment via a non-QND interaction, i.e.,
$[H_S,  H_{SR}]  \ne  0$  such   that  along  with  a  loss  in  phase
information, energy dissipation also takes place.  The evolution has a
Lindblad form which in the interaction picture is given by \cite{bp02}
\begin{eqnarray}
{d \over dt}\rho^s(t) & = & \gamma_0 (N + 1) \left(\sigma_- 
\rho^s(t) \sigma_+ - {1 \over 2}\sigma_+ \sigma_- \rho^s(t) -{1 
\over 2} \rho^s(t) \sigma_+ \sigma_- \right) \nonumber \\ & & + 
\gamma_0 N \left( \sigma_+ \rho^s(t) \sigma_- - {1 \over 
2}\sigma_- \sigma_+ \rho^s(t) -{1 \over 2} \rho^s(t) \sigma_- 
\sigma_+ \right) \nonumber \\ & & - \gamma_0 M \sigma_+ 
\rho^s(t) \sigma_+ -\gamma_0 M^* \sigma_- \rho^s(t) \sigma_- . 
\label{4a6} 
\end{eqnarray}
Here
\begin{equation}
N = N_{\rm th}(\cosh^2 r + \sinh^2 r) + \sinh^2 r, \label{4a7} 
\end{equation}
\begin{equation}
M = -{1 \over 2} \sinh(2r) e^{i\Phi} (2 N_{th} + 1), 
\label{eq:M} 
\end{equation}
and
\begin{equation}
N_{\rm th} = {1 \over e^{\hbar \omega /(k_B T)} - 1}, 
\label{4a9} 
\end{equation}
where $N_{\rm  th}$ is  the Planck distribution  giving the  number of
thermal  photons  at  the  frequency  $\omega$, and  $r$,  $\Phi$  are
squeezing parameters of  the bath. The case of  a thermal bath without
squeezing can be obtained from  the above expressions by setting these
squeezing  parameters  to zero.  $\gamma_0$  is  a constant  typically
denoting the system-environment  coupling strength.  This equation can
be expressed in a manifestly Lindblad form as
\begin{equation}
\frac{d}{dt}\rho^s(t) = \sum_{j=1}^2\left(
2R_j\rho^s R^{\dag}_j - R_j^{\dag}R_j\rho^s - \rho^s R_j^{\dag}R_j\right), 
\label{Lindblad}
\end{equation}
where    $R_1   =    (\gamma_0(N_{\rm   th}+1)/2)^{1/2}R$,    $R_2   =
(\gamma_0N_{\rm  th}/2)^{1/2}R^{\dag}$. Here  $R =  \sigma_-\cosh(r) +
e^{i\Phi}\sigma_+\sinh(r)$,         and         $\sigma_{\pm}        =
\frac{1}{2}\left(\sigma_x  \pm i\sigma_y\right)$.   If $T=0$,  so that
$N_{\rm th}=0$,  then $R_2$ vanishes,  and a single  Lindblad operator
suffices.  The  fact that the above  equation can be  expressed in the
form   (\ref{Lindblad})   guarantees   a   Kraus  or   operator-   sum
representation \cite{nc00}  for the  evolution of the  reduced density
matrix.  It can  be seen that the reduced  density matrix, obtained by
solving  Eq.   (\ref{4a6}) in  the  Bloch  form,  shrinks towards  the
asymptotic equilibrium state $\rho_{asymp}$, given by
\begin{equation}
\rho_{asymp} = \pmatrix{1-p & 0 \cr 0 & p}, \label{4a13}
\end{equation}
where $p = \frac{1}{2}\left[1 + \frac{1}{(2N+1)}\right]$. For the case
of zero squeezing and zero  temperature, this action corresponds to an
amplitude-damping  channel   \cite{nc00,gp}  with  the   Bloch  sphere
shrinking to  a point representing  the state $|0 \rangle$  (the south
pole of  the Bloch sphere) while for  the case of finite  $T$ but zero
squeezing,   the   above   action   corresponds   to   a   generalized
amplitude-damping  channel   \cite{nc00,gp}  with  the   Bloch  sphere
shrinking to  a point  along the  line joining the  south pole  to the
center of the Bloch sphere. The  center of the Bloch sphere is reached
in the limit of infinite  temperature.  Thus, the interaction with the
environment provides a contractive map, such that the asymptotic state
is pure ($p=1$), corresponding to the deletion action \cite{qdele}, or
mixed ($p  < 1$), depending  on environmental conditions.   For finite
$T$  and   bath  squeezing,  the  above  corresponds   to  a  squeezed
generalized amplitude damping channel \cite{sqgen}.

In Eq. (\ref{3p}), the parameter $\alpha$  is given by
\begin{equation}
\alpha = \sqrt{\gamma^2_0 |M|^2 - \omega^2}, \label{3n}
\end{equation}
while 
\begin{equation}
\gamma^{\beta} = \gamma_0 (2 N + 1).
\label{eq:gammabeta}
\end{equation}


\begin{thebibliography}{100}
\bibitem{kraus} K. Kraus, Phys. Rev. D \textbf{35}, (1987) 3070.

\bibitem{mu88} H. Maassen and J. B. M. Uffink, Phys. Rev. Lett.
\textbf{60}, (1988) 1103.

\bibitem{deu83} D. Deutsch, Phys. Rev. Lett. \textbf{50}, (1983) 631.

\bibitem{nc00} M. Nielsen and I. Chuang, \textit{Quantum Computation
and Quantum Information} (Cambridge 2000).

\bibitem{delg} A.  Galindo, M.A. Martin-Delgado, Rev.  Mod. Phys. {\bf
74}, (2000) 347.

\bibitem{pb89} D. T. Pegg and S. M. Barnett, J. Mod. Opt. {\bf
36}, (1989) 7; Phys. Rev. A \textbf{39}, (1989) 1665.

\bibitem{abe} S. Abe, Phys. Lett. A \textbf{166}, (1992) 163.

\bibitem{wiwe} S. Wehner and A. Winter, eprint arXiv:0710.1185.

\bibitem{ii81} I. D. Ivanovic, J. Phys. A \textbf{14}, (1981) 3241.

\bibitem{dur05} T. Durt, J. Phys. A: Math. Gen. \textbf{38}, (2005) 5267.

\bibitem{kl51} S. Kullback and R. A. Leibler, 
Ann. of Math. Stat. \textbf{22}, (1951) 79.

\bibitem{mas07} S. Massar, eprint quant-ph/0703036.

\bibitem{pp98} V. Perinova, A. Luks and J. Perina, \textit{Phase
in Optics} (World Scientific, Singapore 1998).

\bibitem{pad27} P. A. M. Dirac, Proc. R. Soc. Lond. A {\bf
114}, (1927) 243.

\bibitem{sg64} L. Susskind and J. Glogower, Physics \textbf{1}, (1964) 49.

\bibitem{cn68} P. Carruthers and M. M. Nieto,
Rev. Mod. Phys. \textbf{40}, (1968) 411.

\bibitem{ssw90} J. H. Shapiro, S. R. Shepard and N. C. Wong,
Phys. Rev. Lett. \textbf{62}, (1989) 2377.

\bibitem{scr93} \textit{Quantum  Phase and Phase Dependent Measurements},
Eds. W. P. Schleich and S. M. Barnett, Phys. Scr. (Special issue) {\bf
T48}, (1993) 1-144.

\bibitem{ssw91} J. H. Shapiro and S. R. Shepard, Phys. Rev. A
\textbf{43}, (1991) 3795.

\bibitem{mh91} M. J. W. Hall, Quantum Opt. \textbf{3}, (1991) 7.

\bibitem{as92} G. S. Agarwal, S. Chaturvedi, K. Tara and
V. Srinivasan, Phys. Rev. A \textbf{45}, (1992) 4904.

\bibitem{qdele}  R. Srikanth and S. Banerjee,
Phys. Lett. A \textbf{367}, (2007) 295; quant-ph/0611263. 

\bibitem{sr07} S. Banerjee and R. Srikanth, 
Phys. Rev. A:\textbf{76}, (2007) 062109; eprint arXiv:0706.3633.

\bibitem{bp02} H.-P. Breuer and F. Petruccione, \textit{The Theory 
of Open Quantum Systems} (Oxford University Press 2002). 

\bibitem{bg07} S. Banerjee and R. Ghosh, J. Phys.
A: Math. Theo. \textbf{40}, (2007) 13735; eprint quant-ph/0703054.

\bibitem{bvt80} V. B. Braginsky, Yu. I. Vorontsov and 
K. S. Thorne, Science \textbf{209}, (1980) 547. 

\bibitem{bk92} V. B. Braginsky and F. Ya. Khalili, in 
\textit{Quantum Measurements}, edited by K. S. Thorne   
(Cambridge University Press, Cambridge 1992).

\bibitem{wm94} D. F. Walls and G. J. Milburn, \textit{Quantum 
Optics} (Springer, Berlin 1994). 

\bibitem{zu84} W. H. Zurek, in \textit{The Wave-Particle Dualism},  
edited by S. Diner, D. Fargue, G. Lochak and F. Selleri 
(D. Reidel Publishing Company, Dordrecht 1984).

\bibitem{ct80} C .M. Caves, K. D. Thorne, R. W. P. Drever,
V. D. Sandberg and M. Zimmerman, Rev. Mod. Phys. \textbf{52}, (1980) 341.

\bibitem{bo96} M. F. Bocko and R. Onofrio, Rev. Mod. Phys. {\bf
68}, (1996) 755.

\bibitem{sb06} S. Banerjee, J. Ghosh and R. Ghosh,
Phys. Rev. A \textbf{75}, (2007) 062106; eprint quant-ph/0703055.

\bibitem{gp} S. Banerjee and R. Srikanth, 
Euro. Phys. J. D \textbf{46}, (2008) 335; eprint quant-ph/0611161.

\bibitem{sqgen} R. Srikanth and S. Banerjee, 
Phys. Rev. A \textbf{77}, (2008) 012318; arXiv:0707.0059.

\bibitem{as96} G. S. Agarwal and R. P. Singh, Phys. Lett. A
\textbf{217}, (1996) 215.

\bibitem{mr78} M. A. Rashid, J. Math. Phys. \textbf{19}, (1978) 1391.

\bibitem{ap90} G. S. Agarwal and R. R. Puri, Phys. Rev. A \textbf
{41}, (1990) 3782.

\bibitem{at72} F. T. Arecchi, E. Courtens, R. Gilmore and H.
Thomas, Phys. Rev. A \textbf{6}, (1972) 2211.

\bibitem{holevo} A. S. Holevo,
\textit{Probabilistic and Statistical Aspects of Quantum Theory}
(North Holland 1982).

\bibitem{gp93} N Gisin and I. C. Percival, J. Phys. A: 
Math. Gen. \textbf{26}, 2233 (1993).

\bibitem{gis02}  N. Gisin, G. Ribordy, W. Tittel and H. Zbinden,
Rev. Mod. Phys. \textbf{74}, 145 (2002).

\bibitem{cs85} Caves C M and Schumaker B L, Phys. Rev. A 
\textbf {31}, (1985) 3068; Schumaker B L and Caves C M,  
Phys. Rev. A \textbf {31}, (1985) 3093.

\end{thebibliography}
\end{document}